\documentclass[12pt,secnumroman,showpacs,preprintnumbers,amsmath,amssymb]{revtex4}
\usepackage{graphicx,epsfig,epsf}
\usepackage{dcolumn}
\usepackage{bm}
\usepackage{slashed}
\DeclareMathAlphabet{\mathpzc}{OT1}{pzc}{m}{it}

\begin{document}

\title{\Large\sc The Static String}

\author{\sc Sergio Giardino}
 \email{giardino@ime.unicamp.br}
\address{IMECC/Unicamp, Brazil\vspace{1cm}}

\begin{abstract}
In this article the quantum fluctuation of a
rigid and static string is reported to be identical to a free quantum
particle. Solutions similar to this static string have already been found in the
semi-classical quantizaton of pulsating strings, and our results show
that the semi-classical quantization of pulsating strings is, in
some cases, a perturbation of static 
strings. We also interpret the energy of the static string as a lower
bound for the pulsating string and speculate about a description of quantum
mechanics in terms of semi-classical string theory.
\end{abstract}
\noindent 

\maketitle

\section{Introduction}

The semi-classical quantization of a pulsating string in $AdS_5\times S^5$
\cite{Minahan:2002rc,Engquist:2003rn} is executed in such a manner that the string is
allowed to move along the radial dimension $\rho$ of $AdS_5$ and  along the angular coordinates $\theta$ and
$\phi$ of the 
$S^2$ sector of $S^5$. The ansatz $t=\tau$, $\rho=\rho(\tau)$, $\theta=\theta(\tau)$ and
$\phi=m\sigma$ leads to the Nambu-Goto action
\begin{equation}\label{nga}
S=-m\sqrt{\lambda}\int\,dt\,\sin\theta\sqrt{\cosh^2\rho-\dot\rho^2-\dot\theta^2},
\end{equation}
where $\tau$ is the proper time of the string and $\sigma$ is the
length parameter of the string. From this action the canonical
Hamiltonian is obtained, and it allows the wave-function
and the energy spectrum to be calculated. However, the classical equations of motion
mean that $\dot\rho=0$, and thus the string does not move 
along this direction, instead moving in the spherical sector only.
 Although the classical string does not move in $AdS_5$, the simple
 fact that it might move along $\rho$ implies that there is a quantum
 effect associated with this direction. This is due to canonical momentum associated with $\rho$
 obtained from (\ref{nga}), and part of the Hamiltonian operator
 that describes quantum behavior and determines the quantum energy
 spectrum. The quantum effect of this radial degree of freedom is not important
 when considering high quantum numbers. On the other hand, it raises
 the question of a string with no motion along any coordinate 
 that allows obtain a canonical Hamiltonian and consequently
 quantum fluctuations in  its energy.

The possibility of a classical string being used to
 construct a quantum model enables us to establish a relation between
 string theory and quantum mechanics. The quantum harmonic
 oscillator has already been obtained from a
 classical pulsating string \cite{Minahan:2002rc} in  Minkowski
 space-time, and a generalization of this method has obtained a
 variety of quantum models in different space times
 \cite{Giardino:2013sia}. In this article we go further and determine
  that the quantum free particle may be obtained from a classical static string.

On the other hand, quantum fluctuations do not depend only on the
string; it is also determined by the topology and geometry of the space-time
where the string is located. This fact has already been observed in
the semi-classical quantization of pulsating strings in various backgrounds.
\cite{Giardino:2011jy,Arnaudov:2010by,Arnaudov:2010dk,Beccaria:2010zn,Dimov:2004xi,Smedback:1998yn}. We
develop this point further by calculating the differences in the quantum
fluctuations of a static string in a $(2+1)-$dimensional space both
with and without the effect of the other coordinates of the space.
 This idea has already been
explored in the case of pulsating strings and of free falling
strings \cite{Giardino:2013sia}, and in this article we propose to
extend it to static 
strings. The results indicate that the dimension of the space may indeed 
contribute to quantum fluctuations, but the $(2+1)-$dimensional is
interesting as a toy model in order to conceptualize an understanding of the problem.

This article is organized as follows: section (\ref{s2})
describes a classical static string in a $(2+1)-$dimensional space-time
and builds a general quantum model associated with it. In section (\ref{s3}), the
embedding of a world-sheet of a pulsating string and of a free
falling string in a $3-$dimensional plane
space are reviewed and the quantum model of static strings in these
space-times is described using polar coordinates. In section
(\ref{s4}), quantum fluctuations in 
periodic coordinates are considered in spherical and toroidal spaces;
however, the quantum discussion of the toroidal space is qualitative only. Section (\ref{s5}) contains our conclusions.  

\section{the semi-classical description}\label{s2}

We are looking for a static string in a $(2+1)-$dimensional section of
 space-time with the line element
\begin{equation}\label{line}
ds^2=-dt^2+dx^2+g^2dy^2,
\end{equation}
where $g=g(x)$ is a function. The classical motion can be studied by
means of the Virasoro constraint,
\begin{equation}\label{virasoro_0}
-\big(t^{\prime\,2}+\dot t^2 \big)+x^{\prime\,2}+\dot x^2+g^2\big(y^{\prime\,2}+\dot y^2 \big)=0,
\end{equation}
where the dot means a derivative relative to the proper time $\tau$
and the prime means a derivative in relation to the string length parameter
$\sigma$. A static solution can be obtained if  $\dot
x=\dot y=0$ and $r^\prime=t^\prime$, where the simplest choice is, of course,
$r^\prime=t^\prime=0$. It will also be ascertained that $y^\prime$ is
a constant, and then we must choose $y$ as an angular
coordinate. Finally, (\ref{virasoro_0}) implies that $\dot
t^2=g^2y^{\prime\,2}$, with $g$ calculated at the position of the string. Hence the final ansatz for the string is
\begin{equation}\label{ansatz}
t=\kappa\tau,\qquad
x=x(\tau)\qquad\mbox{and}\qquad y=m\sigma + y(\tau),
\end{equation}
where $\kappa$ and $m$ are constants. However, in order to study the
quantum corrections to the classical 
energy, the Nambu-Goto action is used. In the general case where every
coordinate is a function of  $\tau$ and $\sigma$, the Nambu-Goto action is
\begin{equation}\label{nga}
S=-\frac{\sqrt{\lambda}}{2\pi}\int d\tau\int d\sigma\sqrt{\big(\dot
t\, x^{\prime}-\dot x\, t^{\prime}\big)^2+g^2\big(\dot
t\, y^{\prime}-\dot y\, t^{\prime}\big)^2-g^2\big(\dot
y\, x^{\prime}-\dot x\, y^{\prime}\big)^2},
\end{equation}
and, according to the ansatz (\ref{ansatz}), (\ref{nga}) becomes 
\begin{equation}\label{nga2}
S=-\sqrt{\lambda}\,\int d\tau\,g\,y^\prime\sqrt{\dot t^2-\dot x^2}.
\end{equation}
This classical action has no $\dot y$ dependence, which means 
there is no classical momentum associated with this coordinate. 
Consequently, no quantum fluctuation in this direction is observed. The
canonical Hamiltonian in this case is expressed as
\begin{equation}\label{hamil}
H=\dot t\,\sqrt{\Pi^2+\lambda\,y^{\prime\, 2}\,g^2},\qquad\mbox{where}\qquad
\Pi=\sqrt{\lambda}\,y^\prime\,g\,\frac{\dot x}{\dot t^2-\dot x^2}
\end{equation} 
is the canonical momentum for the radial direction. The potential term
of the Hamiltonian is constant, thus we use the square of it in the
Schr\"{o}dinger equation, so that
\begin{equation}\label{schrod}
\big(\hat\Pi^2+\mathpzc{E}^2
\big)\Psi=\frac{\mathcal{E}^2}{\kappa^2}\Psi,
\end{equation}
where $\mathpzc{E}=\sqrt{\lambda}\,\dot t$ and $\dot t=\kappa$ have been used as the classical
energy of the string and $\dot t^2=g^2\,y^{\prime\,2}$ has been used from (\ref{virasoro_0}). Thus, the quantum solutions are free
particles whose squared energy is given by the difference between the squared
classical energy $\mathpzc{E}^2$ and the squared quantum oscillation $\mathcal{E}^2/\kappa^2$, 
\begin{equation}
E^2=\left|\frac{\mathcal{E}^2}{\kappa^2}-\mathpzc{E}^2\right|.
\end{equation}
 The eigenvalue $\pm E^2$ can be either
positive or negative, and hence the  static string has the quantum effect of
generating free oscillating particles and non-oscillating modes in every
point in space. Quantum fluctuation depends on the function
$g$, which determines the geometry of the space and the momentum
operator $\hat\Pi$. Some possibilities for
$g$ have already been discussed in the context of moving strings
\cite{Giardino:2013sia}, and we bring them to
the context of the static string in the next section.

\section{POLAR COORDINATES}\label{s3}

The $(2+1)-$dimensional geometries where the string lies have been
classified according to the function $g$, with $x=r$ a radial
coordinate \cite{Giardino:2013sia}; the results are summarized in
the table below. 
\begin{center}
\begin{tabular}{|l|l|l|}\hline
\rule[-1.5mm]{0mm}{5mm}
$g^2$ & radial rank & topology\\ \hline\hline
\rule[-1.5mm]{0mm}{5mm}$\ell^2r^n\qquad$ & $n=1\qquad r\in[0,\infty)$& non-physical \\
\rule[-1.5mm]{0mm}{5mm} & $n=2\qquad r\in[0,\infty)\qquad $& infinite plane\\
\rule[-1.5mm]{0mm}{5mm} & $n>2\qquad r\in[0,\,\mathcal{R}]\qquad $& finite cone\\
\hline
\rule[-1.5mm]{0mm}{5mm}$\ell^2r^{-n}\qquad$& $n=1\qquad r\in[0,\infty)\qquad $& non-physical\\
\rule[-1.5mm]{0mm}{5mm} & $n>1\qquad r\in[\mathpzc{R},\infty)\qquad $& punctured plane\\
\hline
\end{tabular}
\end{center}
Where $\ell$ is a dimensional constant that gives $g$ the length
dimension for each $n$,
\begin{equation}
\mathcal{R}=\left(\frac{2}{\ell\,n}\right)^{\frac{2}{n-2}}\qquad\mbox{and}\qquad\mathpzc{R}=\left(\frac{\ell\,n}{2}\right)^{\frac{2}{n-2}}.
\end{equation}
The above geometries were used for studying pulsating strings when
$g^2=\ell^2r^n$ and free-falling strings when
$g^2=\ell^2r^{-n}$. Both of these geometries support the static string, however
the quantum fluctuations in each of them are different, therefore we
analyze the cases separately, according to the function $g$. 

\subsection{$g^2=\ell^2r^n$}
In this case, the Schr\"{o}dinger equation
\begin{equation}\label{schrod_pulsating}
\Psi^{\prime\prime}+\frac{n}{2r}\Psi^{\prime}\pm E^2\Psi=0.
\end{equation}
has different solutions for each sign above, and they are expressed in terms of Bessel functions 
\begin{eqnarray}\label{free_puls}
&&\Psi_+=\frac{1}{r^{\frac{n-2}{4}}}\Big[A\,J_{\frac{n-2}{4}}(E\,r)+B\,Y_{\frac{n-2}{4}}(\mathpzc{E}\,r)\Big]\\
&&\Psi_-=\frac{1}{r^{\frac{n-2}{4}}}\Big[C\,I_{\frac{n-2}{4}}(E\,r)+D\,K_{\frac{n-2}{4}}(\mathpzc{E}\,r)\Big],
\end{eqnarray}
where the $A$, $B$, $C$ and $D$ are integration constants. The
wave-function $\Psi_+$ describes oscillating modes so that
$\mathcal{E}^2/\kappa^2>\mathpzc{E}^2$, and $\Psi_-$ obviously corresponds to
non-oscillating modes so that $\mathcal{E}^2/\kappa^2<\mathpzc{E}^2$.  The
quantum fluctuation energy spectrum
can thus be expressed for each case as
\begin{equation}
\mathcal{E}_+^2=\kappa^2\big(\mathpzc{E}^2+E^2\big)\qquad\mbox{and}\qquad
\mathcal{E}_-^2=\kappa^2\big(\mathpzc{E}^2-E^2\big), 
\end{equation}
where the discretization or the continousness of the energy spectrum
depends on the eigenvalue $E^2$. It can also be seen that 
$\kappa^2\mathpzc{E}^2$ is the zero point squared energy, so that in the case of  $\mathcal{E}_+^2$
the classical energy is a lower bound and in the case of the
$\mathcal{E}_-^2$ the classical energy is an upper bound. 
 
The free particle wave-functions are normalizable for the finite
spaces where $n>2$. However, 
as the Bessel functions $Y$ and $K$ are singular at $r=0$, we make
$B=0$ and $D=0$. The Bessel function $I$
goes to infinity if $r\to\infty$, hence for $n=2$, where
$r\in[0,\,\infty)$, $C=0$ as well, and thus non-oscillating solutions
occur only for $n>2$. However, in the $n=2$ case the wave-function is
not normalizable. This 
problem can be circumvented when the solutions are localizable
\cite{Giardino:2013sia}, something that is obtained if the
wave-function obey
\begin{equation}\label{dirac}
\intop_{0}^{\infty} \Psi\,\Psi^\star\sqrt{-\,G}\,dr=\delta(r),
\end{equation}
where $G$ is the determinant of the metric tensor. The Dirac delta
function expressed in terms of Bessel functions in a
$(d+1)-$dimensional space is given as
\begin{equation}\label{wave_delta}
\delta^{d+1}\big(\epsilon^2-\eta^2\big)=\intop_0^\infty dr\,r\,J_\mu(\epsilon\,r)\,J_\mu(\eta\,r),
\end{equation}
and as the wave-function for $n=2$ satisfies (\ref{wave_delta}), it
can indeed be understood as a free particle.
 The non-oscillating solutions are either
non-localizabe or non-normalizable for $n=2$, and thus they have been
disregarded. On the other hand, for $n>2$, $r$ is finite 
and $\Psi_-$ may be admitted. 
Non-oscillating  modes are known from the tunneling phenomenon, where they
describe classically prohibited regions. They are not
localizable, and certainly cannot be understood as particles.

The energy of oscillating modes can either be discrete or
continuous, and the energy spectra of 
non-oscillating solutions are always continuous. For $n=2$ the energies
are always continuous, and for $n>2$ continuous and discrete energies
coexist. The Bessel function $J$ for $n>2$ determines
quantum solutions if a root occurs at the
border $r=\mathcal{R}$ of the space. Then there are quantized and non-quantized
solutions, so that the eigenvalues for the quantized solutions obey
\begin{equation}\label{qenergy}
E_N=\frac{R^{(N)}}{\mathcal{R}},
\end{equation}
so that $N\in\mathbb{N}$ and $R^{(N)}$ is a root of $J$. As
$\mathcal{R}$ and $R^{(N)}$ are fixed for each $N$, there is just one value of
$E$ which obeys (\ref{qenergy}), and then this class of
solutions is quantized.

\subsection{$g^2=\ell^2r^{-n}$}
This case is very similar to the preceding one, and the Schr\"{o}dinger equation
\begin{equation}\label{schrod_pulsating}
\Psi^{\prime\prime}-\frac{n}{2r}\Psi^{\prime}\pm E^2\Psi=0
\end{equation}
also has its solutions given in terms of Bessel functions
\begin{eqnarray}\label{free_puls_2}
&&\Psi_+=r^{\frac{n+2}{4}}\Big[A\,J_{\frac{n+2}{4}}(E\,r)+B\,Y_{\frac{n+2}{4}}(E\,r)\Big]\\
&&\Psi_+=r^{\frac{n+2}{4}}\Big[C\,I_{\frac{n+2}{4}}(E\,r)+D\,K_{\frac{n+2}{4}}(E\,r)\Big],
\end{eqnarray}
where $A$, $B$, $C$ and $D$ are integration constants. $\Psi_+$ is
non-normalizable but is localizable in the sense of (\ref{dirac}) for
either $B=0$ or $A=0$. $\Psi_-$ is evanescent and normalizable for $C=0$. 

There are discrete quantum solutions for $\Psi_+$ when a root  of the Bessel
function $J$ or $Y$ occurs at the border of the space, where
$r=\mathpzc{R}$, and the energies for the discrete quantized solutions obey
\begin{equation}\label{energy2}
E_N=\frac{R^{(N)}_{\mathcal{Z}}}{\mathpzc{R}},\qquad \mbox{so that}\qquad
N\in\mathbb{N}\qquad\mbox{and}\qquad\mathcal{Z}=\{J,\,Y\},
\end{equation}
where $N\in\mathbb{N}$ and $R^{(N)}_{\mathcal{Z}}$ is a root of
either $J$ or $Y$. $\Psi_+$ allows the Bessel function $Y$
because the singular point $r=0$ is out of the space. Then, there are more
allowed quantum wave-functions than in the preceding case. 

These solutions, for both of $g^2$, enable us to state that the
oscillating quantum fluctuations of the static string are free
particles. Thus, the quantum problems of 
the harmonic oscillator and the free particle can be described in
terms of the semi-classical approximation of string theory. In this
sense, we can speculate that there is a correspondence between
semi-classical string theory and quantum mechanics. We note in these cases
the usefulness of string 
theory as an instrument to describe different physical
theories where there is no string at all, something that has already
been developed in 
the context of $AdS/CFT$ correspondence. If this hypothesis is
correct, the role of string theory in physics remains an open
question: it can either be a mathematical framework or a fundamental
theory in which strings are a physical reality.

\section{ angular coordinates}\label{s4}

With space coordinates angular and
 periodic in (\ref{nga2}), different topologies can be studied,
 and we consider here the sphere and the
torus. The classical string on a sphere has been studied in the context
of a pulsating string \cite{Minahan:2002rc}, and we extend it to the
static case. The string on a torus is a new possibility, and we are
interested in pointing out the difference of the quantum
fluctuations due to the topology change in relation to the spherical case.

\subsection{STRING ON A SPHERE}
The sphere is obtained for $x=\theta$, $y=\varphi$ and $g=\sin\theta$. The static
string ansatz for this geometry is
\begin{equation}
t=\kappa\tau,\qquad \theta=\theta(\tau)\qquad\mbox{and}\qquad \varphi=m\sigma,
\end{equation}
and the Nambu-Goto action, the Hamiltonian and the Schr\"{o}dinger
equation comes straight from (\ref{schrod}). However,
contrary to the former cases, the $\hat\Pi^2$ operator changes
with the dimension of space-time because the Laplacian operator
changes with the determinant of the metric. If we use a $(2+1)-$dimensional
space, we obtain
\begin{equation}\label{legendre}
\frac{1}{\sin\theta}\frac{d}{d\theta}\Big(\sin\theta\frac{d}{d\theta}\Psi\Big)\pm E^2\Psi=0,
\end{equation}
and if a ten dimensional space is used, the result is
 \begin{equation}\label{jacobi}
\frac{1}{\sin^3\theta\cos\theta}\frac{d}{d\theta}\Big(\sin^3\theta\cos\theta\frac{d}{d\theta}\Psi\Big)\pm E^2\Psi=0.
\end{equation}
The solutions for (\ref{legendre}) and (\ref{jacobi}) with the plus
sign have already been 
used to semi-classically quantize the pulsating string in 
Lunin-Maldacena space-time \cite{Giardino:2011jy}, respectively in terms of
Legendre and Jacobi polynomials, so that
\begin{equation}
\Psi_{Legendre}=\sqrt{2N+1}P_N(\cos\theta)\qquad\mbox{and}\qquad \Psi_{Jacobi}=2\sqrt{N+1}P_N^{(0,1)}(1-\cos^2\theta),
\end{equation}
where $N\in\mathbb{N}$. They are oscillating functions whose
eigenvalues are respectively
\begin{equation}\label{energy_spectra}
E_{Legendre}^2=N(N+1)\qquad\mbox{and}\qquad E_{Jacobi}^2=4N(N+2).
\end{equation}
On the other hand, the $-E^2$ case in (\ref{legendre}) is written in
terms of Legendre functions with complex index, the conical function
$P_{-\frac{1}{2}+i\tau}$, where $\tau=\sqrt{4E^2-1}$ and the energy is not
discrete in this case. These functions are normalizable, and then they
can
be used as wave-functions for the non-oscillating modes. On the other
hand, the solution for (\ref{jacobi}) with $-E^2$ is hypergeometric
function $F\big(1-q,\,1+q;1,1;\,\frac{1-\cos\theta}{2}\big)$, with
$q=\sqrt{E^2-1}$. This function  generates a divergent hypergeometric series and
 consequently cannot be a normalizable wave-function. 

The results show that the former pulsating string semi-classical
quantization results have indeed been obtained from perturbation
of the static string wave-function, and then this is a more
fundamental situation. The results also demonstrate that the effect of other
dimensions of space are in fact sensible, which leads to very
different energy spectra (\ref{energy_spectra}) and also determines whether
the non-oscillating modes exist or not.

\subsection{STRING ON A TORUS}
The embedding coordinates of the torus in a three-dimensional space are
\begin{equation}\label{emb_tor}
x^1=\big(R+\rho\cos\theta\big)\cos\phi,\qquad
x^2=\big(R+\rho\cos\theta\big)\sin\phi,\qquad\mbox{and}\qquad x^3=\rho\sin\theta,
\end{equation}
where $R$ and $\rho$ are the radii of the torus so that $R>\rho$. The
range of both of the angular coordinates is $[0,\,2\pi]$,
$\theta$ is the angular coordinate using $\rho$ as radius and $\phi$
is the angular coordinate using $R$ as its radius. From
(\ref{emb_tor}), the metric of the
toroidal $(2+1)-$dimensional space-time is
\begin{equation}
ds^2=-dt^2+\rho^2d\theta^2+\big(R+\rho\cos\theta\big)^2d\phi^2.
\end{equation}
Using the usual ansatz
\begin{equation}
t=\kappa\tau\qquad\theta=\theta(\tau)\qquad\mbox{and}\qquad\phi=m\sigma,
\end{equation}
the Virasoro constraint
\begin{equation}
-\kappa^2+\rho^2\,\dot{\theta}^2+\big(R+\rho\cos\theta\big)^2\phi^{\prime\,2}=0
\end{equation}
is satisfied for $\dot\theta=0$, and then the string is static. Using
$x=\rho\,\theta$, $y=\phi$ and 
$g=R+\rho\cos\theta$ in the Nambu-Goto action (\ref{nga2}), the
Schr\"{o}dinger equation for the free particle follows
 with the determinant of the metric
$\sqrt{-G}=R+\rho\cos\theta$. Changing the variable so that
$\mathpzc{x}=\frac{1+\cos\theta}{2}$, we obtain a Heun-type
equation
\begin{equation}\label{heun}
\Psi_{\mathpzc{x}\mathpzc{x}}+\left[\frac{1}{2}\left(\frac{1}{\mathpzc{x}}-\frac{1}{\mathpzc{1-x}}\right)+\frac{1}{\mathpzc{x}+\frac{R-\rho}{2\rho}}\right]\Psi_{\mathpzc{x}}\pm\frac{E^2}{\mathpzc{x}(1-\mathpzc{x})}\Psi=0,
\end{equation}
where the index $\mathpzc{x}$ means a derivative with respect to this
coordinate and (\ref{heun}) is a Heun-type equation. The solution for
$+E^2$ is periodic, oscillating and orthogonal 
function with appropriate choice of the parameters.  When the eigenvalue is $-E^2$, the periodic and oscillating
character disappear, and then the static string in the torus also has 
the qualitative features that were found in the static string on the
sphere. The toroidal case differs from the spherical case in topology
only. This geometrical difference is hence responsible for the change in the
wave-function and in the energy spectrum, and is then another example 
of the influence of space-time on the quantum behavior of a
string. The discussion of the quantum fluctuations of the static string
in the toroidal space is qualitative because a detailed treatment of the quantum solutions of (\ref{heun}) is
complicated enough to deserve an independent study.

\section{CONCLUSION}\label{s5}
In this paper we introduced a classical circular string that
remains at a definite position in the space. This object can be
semi-classically quantized in various space-times, and a quantum free
particle is then obtained in each case. The difference in relation to the usual
quantum free particle is the split between oscillating and
non-oscillating modes. The oscillating modes are the usual quantum free
particles and their energies are higher than the classical energies of
static strings, while the non-oscillating modes have lower energy than the
classical energy of the static string.  The results also demonstrate
that static string wave-functions have already been applied in 
the semi-classical quantization of pulsating strings. In
fact, the known quantum pulsating string in a sphere is a
static string perturbation, and hence the static string it is possibly
a more fundamental object than a pulsating string.

Furthermore, the results give rise to some fundamental questions about the nature of
string theory. The quantum harmonic oscillator and the quantum free particle can be
obtained semi-classically from string theory, and it is important
to comprehend to what extent quantum mechanics may be reconstructed
from string theory. If every result from quantum mechanics can be
recovered, then string theory is either a more fundamental theory or a
very powerful mathematical framework. However, string theory may
describe only some quantum results, and so the theories are in
fact distinct frameworks with some overlap in specific limits. The
extension of this overlap between string theory and quantum mechanics
seems to be, from the author's standpoint, a very
important question that needs to be answered in order to understand the physical
content of these theories.

\section*{Acknowledgements}
Sergio Giardino is grateful for the financial support of Capes and for
the facilities provided by the IFUSP.  
%
%
%
%

%
%
%
%

\begin{thebibliography}{1}

\bibitem{Minahan:2002rc}
J.~A. Minahan.
\newblock {``Circular semiclassical string solutions on $AdS_5 \times S^5$''}.
\newblock {\em Nucl. Phys.}, {\bf B648}:203--214, (2003) hep-th/0209047.

\bibitem{Engquist:2003rn}
J.~Engquist; J. A.Minahan;~K. Zarembo.
\newblock {``Yang-Mills duals for semiclassical strings on $AdS_5\times S^5$}.
\newblock {\em JHEP}, {\bf 0311}:063, (2003) hep-th/0310188.

\bibitem{Giardino:2013sia}
S.~Giardino.
\newblock {Semi-classical strings in $(2+1)-$dimensional backgrounds}.
\newblock (2013) arXiv:1303.7167[hep-th].

\bibitem{Giardino:2011jy}
Sergio Giardino and Victor Rivelles.
\newblock {Pulsating Strings in Lunin-Maldacena Backgrounds}.
\newblock {\em JHEP}, 1107:057, (2011) arXiv:1105.1353[hep-th].

\bibitem{Arnaudov:2010by}
D.~Arnaudov; H. Dimov; R.~C. Rashkov.
\newblock {On the pulsating strings in $AdS_5 x T^{1,1}$}.
\newblock {\em J.Phys.}, {\bf A44}:495401, (2011) arXiv:1006.1539[hep-th].

\bibitem{Arnaudov:2010dk}
D.~Arnaudov; H. Dimov; R.~C. Rashkov.
\newblock {``On the pulsating strings in Sasaki-Einstein spaces''}.
\newblock {\em AIP Conf.Proc.}, {\bf 1301}:51--58, (2010)
  arXiv:1007.3364[hep-th].

\bibitem{Beccaria:2010zn}
M.~Beccaria; G. V. Dunne; G. Macorini; A. Tirziu; A.~A. Tseytlin.
\newblock {``Exact computation of one-loop correction to energy of pulsating
  strings in $AdS_5 x S^5$''}.
\newblock {\em J.Phys.}, {\bf A44}:015404, (2011) arXiv:1009.2318[hep-th].

\bibitem{Dimov:2004xi}
H.~Dimov; R.~C. Rashkov.
\newblock {``Generalized pulsating strings''}.
\newblock {\em JHEP}, {\bf 0405}:068, (2004) hep-th/0404012.

\bibitem{Smedback:1998yn}
M.~Smedback.
\newblock {``Pulsating strings on $AdS_5\times S^5$''}.
\newblock {\em JHEP}, {\bf 0407}:004, (2004) hep-th/0405102.

\end{thebibliography}
\end{document}